\documentclass[10pt]{article}
\usepackage{latexsym}
\usepackage{amsfonts}
\usepackage{graphicx}
\usepackage[numbers]{natbib}
\usepackage{subfig}
\usepackage{tikz}
\usepackage{multicol}
\usetikzlibrary{shapes.symbols,shapes.geometric,shadows,arrows.meta}
\tikzset{>={Latex[width=1.5mm,length=2mm]}}
\usepackage{flowchart}\usepackage[paperheight=11.69in,paperwidth=8.26in,left=0.61in,right=0.61in,top=0.61in,bottom=0.61in,headheight=1in]{geometry}
\usepackage[utf8]{inputenc}
\usepackage[T1]{fontenc}
\usepackage{helvet}

\usepackage[font=small,labelfont=bf]{caption}

\newenvironment{Figure}
  {\par\medskip\noindent\minipage{\linewidth}}
  {\endminipage\par\medskip}

\usepackage{tikz}
    \usetikzlibrary{shadows}
\usepackage{tcolorbox}
    \tcbuselibrary{skins}

\definecolor{esoblue}{HTML}{0070C0}
\newcommand{\mytitle}[1]{
    \node[fill=white,
        draw=esoblue,
        line width=0.5pt,
        text width=1.75cm,
        inner sep=8pt,
        xshift=-3.7cm]
    at (frame.north){\bfseries\textcolor{black}{#1}};
}

\newtcolorbox{mybox}[2][]{
    enhanced,
    overlay={\mytitle{#2}},
    borderline={.7pt}{0mm}{esoblue},
    frame hidden,
    arc=0mm,
    sidebyside,
    lefthand width=2.5cm,
    segmentation hidden,
    top=15pt,
    #1
}

\newcommand{\arttitle}[1]{\fontsize{24pt}{32pt}\selectfont \textcolor[HTML]{0070C0}{#1}\par}
\newcommand{\artauth}[1]{\fontsize{12pt}{18pt}\selectfont \textbf{#1}\par}
\newcommand{\artaff}[1]{\fontsize{11pt}{16pt}\selectfont #1 \par}

\newcommand{\TheTitle}{The most giant radio structures in the Universe}
\newcommand{\MyName}{Andrea Botteon}
\newcommand{\MyInst}{Leiden Observatory, Leiden University, Niels Bohrweg 2, 2300 RA Leiden, The Netherlands}

\begin{document}

\begin{center}
\arttitle{\TheTitle}\par
\artauth{\MyName}\par
\artaff{\MyInst}
\end{center}\par

\addcontentsline{toc}{section}{{\MyName} - {\it \TheTitle}}
\vspace{0.5cm}
\begin{multicols}{2}


\noindent
Galaxy clusters are the largest and most massive gravitationally bound systems in the Universe, reaching linear sizes of a few Mpc and masses up to $10^{15}$ M$_\odot$. Although the richest clusters may contain thousands of galaxies in their volume, galaxies account only for $\sim$5\% of the cluster total mass which is instead dominated for about $\sim$80\% by dark matter. The remaining $\sim$15\% is in form of intra-cluster medium (ICM), a tenuous ($n_e \sim 10^{-3}-10^{-4}$ cm$^{-3}$) and hot ($T \sim 10^7 - 10^8$ K) plasma whose thermal bremsstrahlung emission is detectable in the X-rays. 

Clusters of galaxies form hierarchically via energetic merger events. During these cosmic collisions, shocks and turbulence are injected in the ICM, and often generate cluster-wide synchrotron sources with steep spectrum ($\alpha > 1$, with $S_\nu \propto \nu^{-\alpha}$, where $S_\nu$ is the flux density at frequency $\nu$ and $\alpha$ is the spectral index). These sources are called radio halos and relics, and are currently observed in a fraction of massive clusters \cite{vanweeren19rev}. The existence of diffuse Mpc-scale synchrotron emission in clusters challenges our understanding of cluster astrophysics as it probes a complex hierarchy of (novel) mechanisms in the ICM that are essentially able to turn gravitational energy into particle acceleration and magnetic field amplification \cite{brunetti14rev}. It is currently thought that radio halos trace turbulent regions where relativistic particles are trapped and re-accelerated through scattering with magnetohydrodynamics turbulence. Instead, radio relics originate as a consequence of the particle acceleration and magnetic field amplification undergoing at merger shocks located in cluster outskirts.

Despite the observed connection between mergers and radio halos and relics, little is known about the generation of synchrotron emission in the very early phase of a merger (Figure~\ref{fig:scheme}). Systems in this merging phase are referred to as pre-merging clusters. Until recently, it was still unclear if a significant fraction of the energy of gas dynamics can be channelled into non-thermal components during the pre-merger phase becoming detectable in the radio band. 

\begin{Figure}
    \centering
    \includegraphics[width=\linewidth]{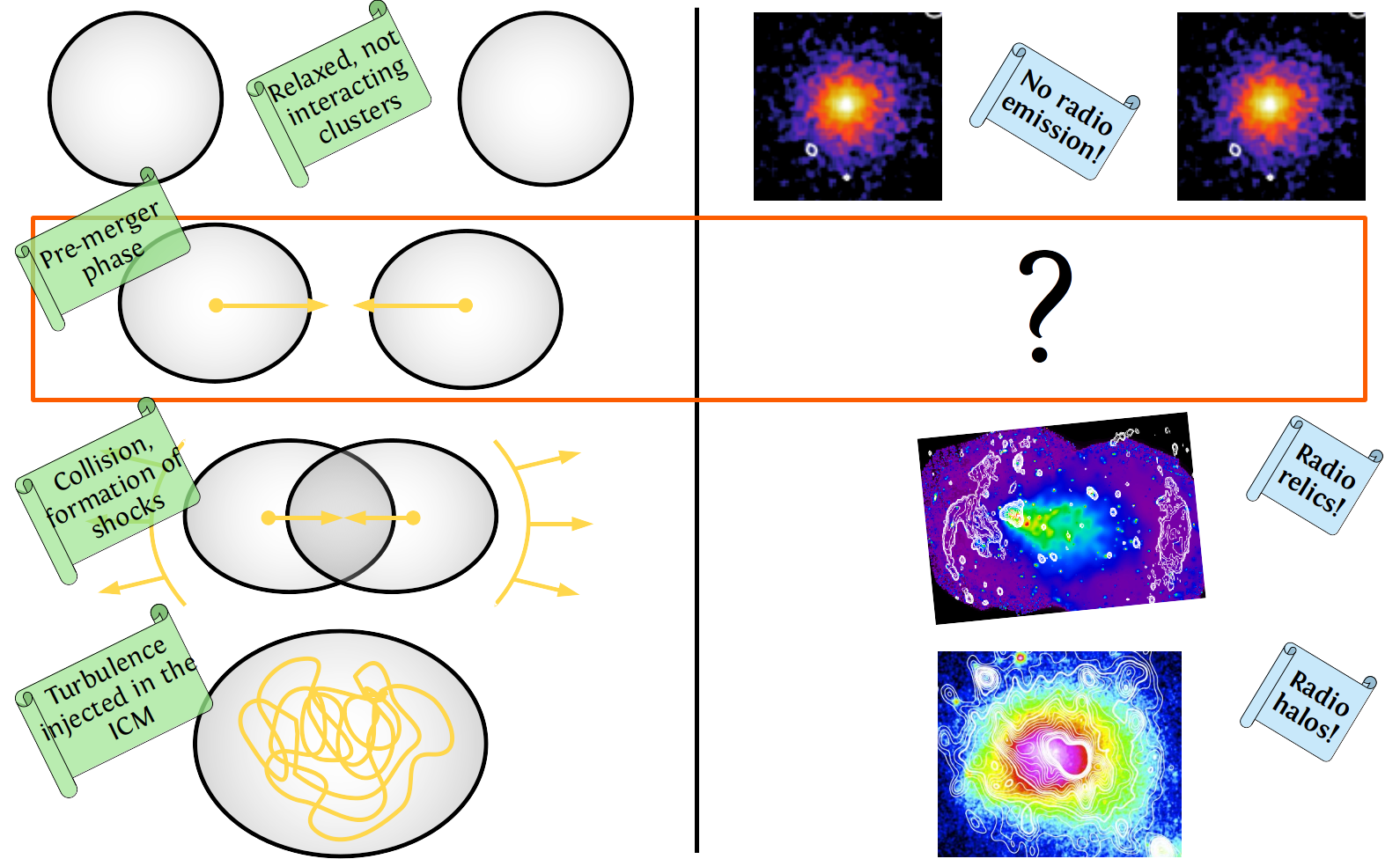}  
    \captionof{figure}{\small Schematic view of a binary cluster merger.}  
    \label{fig:scheme}
\end{Figure}
\vspace{0.5cm}

The ongoing progress in radio instrumentation has recently made possible to enter into a new regime in which non-thermal phenomena can be studied in large-scale structures. In particular, the LOw Frequency ARray (LOFAR) enabled the first detailed observations of galaxy clusters at frequencies of $<$200 MHz thanks to the unprecedented high sensitivity and high resolution in its operational frequency range.

\vspace{0.25cm}
{\fontsize{10pt}{10.8pt}\selectfont \textcolor[HTML]{0070C0}{The discovery of radio bridges}\par}
\noindent 
Pairs of galaxy clusters are rare systems composed of two close-by clusters that are bound together by gravity. Their observation gives us the chance to study the phases of the merger closer to the epoch of the collision and to understand the impact of mergers on cluster evolution. In particular, Abell 1758 is a system located at $z = 0.279$ composed of two massive galaxy clusters separated by a projected distance of $\sim$2 Mpc: Abell 1758N (in the north, the most massive one) and Abell 1758S (in the south). X-ray observations suggest that the two clusters are gravitationally bound but have not interacted yet, that is, they are in a pre-merger phase. In addition, complex cluster dynamics and multiple sub-substructures are observed both in Abell 1758N and Abell 1758S, indicating that each of the two clusters is undergoing its own merger. In \citet{botteon18}, we used an 8~h observation at 144 MHz from the LOFAR Two-meter Sky Survey (LoTSS) to study the well-known radio halo in Abell 1758N and also discovered a new radio halo and a candidate radio relic in Abell 1758S. More importantly, at low resolution we found a hint ($2\sigma$) of a bridge of radio emission connecting the two clusters which required a further study with more sensitive observations. With the deep and multi-frequency follow-up observations of the cluster we were able to firmly claim the presence of a giant radio bridge at 144 MHz that fills the volume between the two clusters (Figure~\ref{BotteonFig2}, left), while only hints of radio emission were observed at 53 and 383 MHz \cite{botteon20}.

\begin{Figure}
    \centering
    \includegraphics[width=\linewidth]{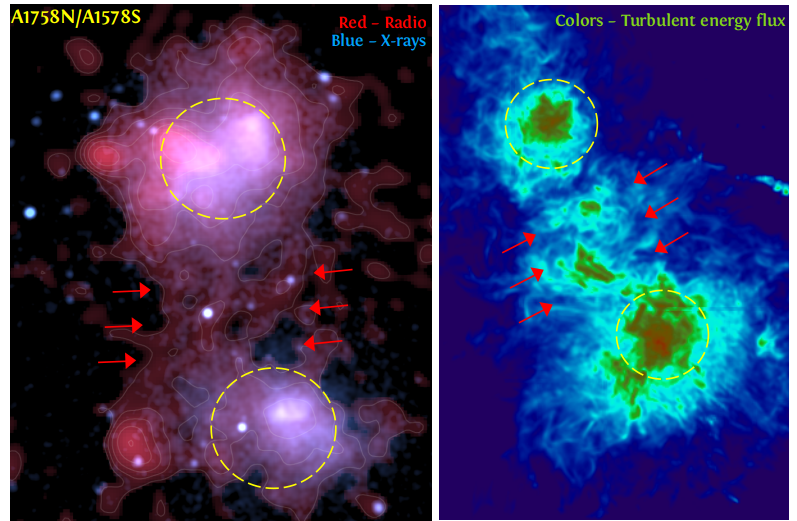}  
    \captionof{figure}{\small Radio bridges in Abell 1758N-1758S (left, \cite{botteon18, botteon20}) and in simulations (right, \cite{brunetti20}). Dashed yellow circles have a diameter of $\sim$1 Mpc.}  
    \label{BotteonFig2}
\end{Figure}
\vspace{0.5cm}

The detection of radio emission from the region between pre-merging galaxy clusters demonstrates, for the first time, that even in environments where the average particle energy is much lower than within a cluster, particles can be (re)accelerated by shocks and/or turbulence at meaningful levels. These regions are in fact at the crossroads between the denser and hotter ICM regions and the colder and rarefied cosmic web and thus allow us to study particle acceleration mechanisms in new regimes, still poorly constrained from theory.

Presently, the pre-merging cluster pairs Abell 1758N-1758S (at $z = 0.279$) and Abell 399-401 (at $z=0.07$, \cite{govoni19}) are the only two cases where a bridge of radio emission between two clusters has been observed. The two systems show remarkable similarities. First of all, each of the two main components of the pairs is a massive cluster, with $M_{500} > 5 \times 10^{14}$ M$_\odot$. Secondly, both are pairs of dynamically disturbed clusters, with all four clusters undergoing mergers and hosting a radio halo. Thirdly, the diffuse radio emission from the bridges connecting the two pairs (detected in both cases with LOFAR 144 MHz observations) spans a scale of $\sim$2-3 Mpc and has similar mean radio emissivity.

Two models have been proposed to explain the acceleration of relativistic electrons and amplification of magnetic fields in the compressed region between clusters: one involves the presence of weak shocks \cite{govoni19}, the other of turbulent motions \cite{brunetti20}. According to the latter, turbulence may produce steep spectrum ($\alpha > 1.3$) and volume filling synchrotron emission in the entire bridge region (Figure~\ref{BotteonFig2}, right). The analysis of radio and X-ray observations of Abell 1758N-1758S are in agreement with these predictions \cite{botteon20}. However, deeper data on this system as well as observations of other cluster pairs are still required to firmly determine the origin of radio bridges.

\vspace{0.25cm}
{\fontsize{10pt}{10.8pt}\selectfont \textcolor[HTML]{0070C0}{LOFAR observations of other cluster pairs}\par}
\noindent 
LOFAR observations at 144 MHz have been used to search for radio emission between clusters in other complex systems with multiple components. For example, in the spectacular cluster chain Abell 781 we did not find any diffuse emission \cite{botteon19a781}, while in the pair Abell 1430 we found an atypical diffuse radio emission in a low-density environment that was dubbed ``Pillow'' \cite{hoeft21}. The Lyra complex is another interesting pre-merging cluster pair that we observed recently \cite{botteon19lyra}. Despite the absence of a radio bridge between the two clusters, we discovered a new radio halo in the main cluster component (Figure~\ref{BotteonFig3}). This halo has a low-surface brightness extension towards the SW, leading to a maximum linear extent of the diffuse radio emission up to $\sim$1.8 Mpc. As for the case of radio bridges, the presence of an extension of a radio halo probes the existence of non-thermal components at large distances from the cluster center. In particular, we speculated that the emission in this system is a consequence of the energy dissipated on small scales due to the interaction between the main cluster of the pair and another galaxy group present in the complex (Figure~\ref{BotteonFig3}).

Compared to the radio bridge pairs Abell 1758N-1758S and Abell 399-401, the other systems observed with LOFAR are less massive, possibly suggesting a role of the mass in the generation of observable levels of synchrotron emission between pre-merging clusters.

\begin{Figure}
    \centering
    \includegraphics[width=\linewidth]{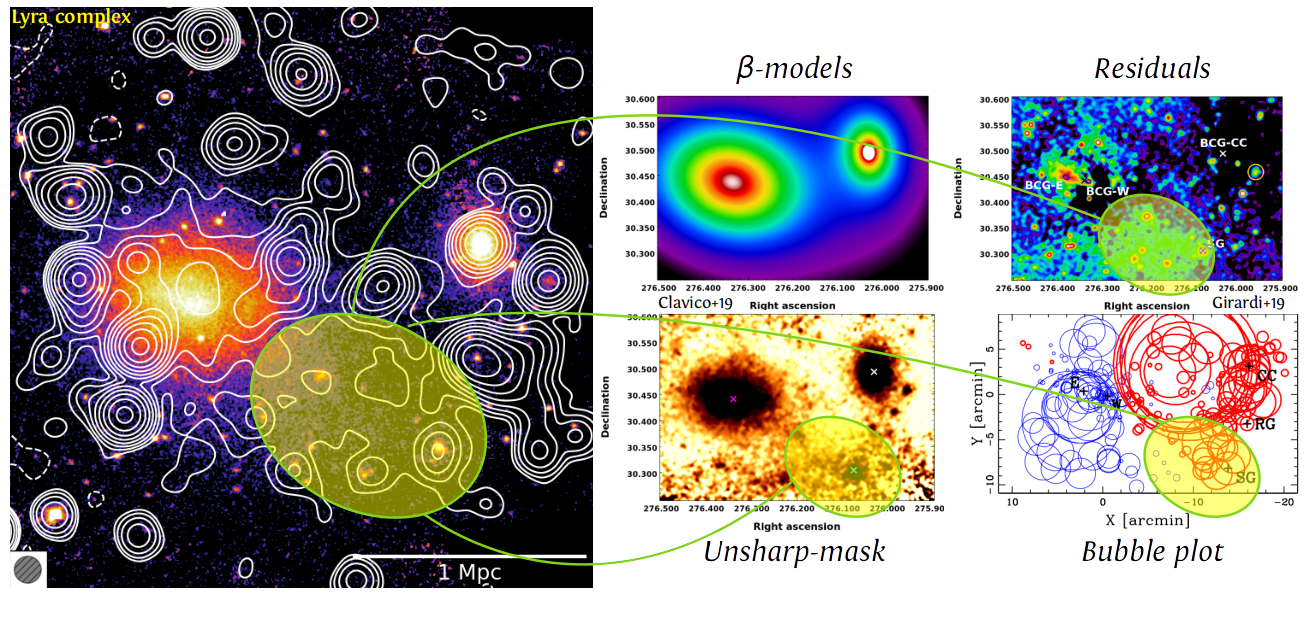}  
    \captionof{figure}{\small LOFAR (contour) and XMM-Newton (color) overlay of the Lyra complex (left). The SW low-surface brightness extension of the newly discovered radio halo is highlighted in the figure. Optical and X-ray observations indicate the presence of substructure in this region (right).}  
    \label{BotteonFig3}
\end{Figure}
\vspace{0.5cm}


\vspace{0.25cm}
{\fontsize{10pt}{10.8pt}\selectfont \textcolor[HTML]{0070C0}{Conclusions}\par}
\noindent
The detection of radio bridges and extensions of radio halos demonstrates the existence of magnetic fields and particle acceleration mechanisms at large distance from the cluster center. Particularly, the discovery of radio bridges in pre-merging cluster pairs probed for the first time that non-thermal phenomena can be generated by mechanisms that are not necessarily related to the energy dissipated as a consequence of cluster mergers. Future radio (e.g. LOFAR2.0 and SKA) and X-ray (e.g. Athena) instruments will play a crucial role to efficiently study cluster outskirts and determine how common are gigantic radio sources in clusters, what are the physical processes that generate them, and what is their connection with the large-scale structure of the Universe.


\vspace{0.35cm}
{\fontsize{9pt}{9.8pt}
\selectfont \textcolor[HTML]{0070C0}{References}\par 

\begingroup
\renewcommand{\section}[2]{}%

\endgroup
}
\end{multicols}

\vspace*{\fill}
\begin{mybox}{Short CV}
    \includegraphics[scale=.3]{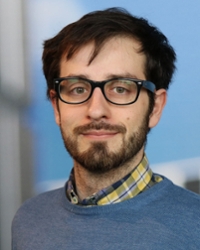}   
    \tcblower
    \begin{tabular}{l@{\hspace{1\tabcolsep}}l}
    2013:& Bachelor in Astronomy, University of Bologna, Italy\\
    2015:& Master in Astrophysics and Cosmology, University of Bologna, Italy\\
    2018:& PhD in Astrophysics, University of Bologna and INAF-IRA, Italy\\
    2018--2019:& Postdoc Researcher, University of Bologna and INAF-IRA, Italy\\    
    2019--present:& Postdoc Researcher, University of Leiden, The Netherlands\\  
    \end{tabular}
\end{mybox}

\end{document}